\documentstyle{aipproc}

\begin{document}
\title{Entropy Production and\\
Heat Generation in\\
Computational Processes}

\author{Nobuko Fuchikami$^*$\thanks{Email: fuchi@phys.metro-u.ac.jp}, 
Hijiri Iwata$^*$ and 
Shunya Ishioka$^{\dagger}$}
\address{$^*$Department of Physics, Tokyo Metropolitan
University\\
Minami-Ohsawa, Hachioji, Tokyo 192-0397, Japan\\
$^{\dagger}$Department of Information Science, Kanagawa University\\
Hiratsuka, Kanagawa 259-1293, Japan}

\maketitle

\vspace*{10mm}

{\small
\noindent
{\bf Summary.} ~To make clear several issues relating with the thermodynamics
of computations, we perform a simulation of a binary device using a
Langevin equation.
Based on our numerical results, we consider how to estimate thermodynamic 
entropy of computational devices. We then argue against the existence of the
so-called residual entropy in frozen systems such as ice.
}

\section*{1. Introduction}

How much energy must be used in a computation?
Landauer\cite{l61} answered to this question in 1961 as follows: 

``Logical irreversibility implies
physical irreversibility'' (p.186 in Ref. [1]) and
requires a finite amount of heat generation.
A data-erasing operation is generally irreversible and induces the heat
generation, which corresponds to a cost of energy, $kT \ln 2$/bit (at 
least), while a reversible operation such as copying can be free
of any cost of energy.

It seems that Landauer's theory has been accepted widely since then
\cite{b97,f97}.
Recently, however, Goto et al. disputed it and claimed that it is possible 
to erase information without producing heat generation by using Quantum Flux 
Parametron 
devices\cite{gylh89}.
The controversy has lead us to a recognition of several problems which
relate to  
the thermodynamics of computations and are still left unclear:

During the computational processes,

q1. when does the heat generation occur?

q2. when is the energy consumed?

q3. when does the entropy increase (and when does it decrease)?

q4. how should the thermodynamic entropy of computational devices be defined?

\section*{2. Data-erasing operation
}
Let us consider a binary device which consists of a bistable potential well 
and a particle in it. When the particle is in the left (right)-hand well, we
assign state ``0'' (``1'') to the device.
Following Landauer, we define a data-erasing operation, RTO (Restore To
One) which resets the device on state ``1'' regardless of the initial state.
In order to erase the information, 
one must carry out the operation RTO without
knowing whether the device is initially in state ``0'' or ``1''. This can 
be done by the following processes:
\begin{equation}
\begin{array}{ccccccccc}
  U_2        & \rightarrow & U_1       & \rightarrow & U_1 + U_3 
             & \rightarrow & U_2 + U_3 & \rightarrow & U_2,\\
  (\rm{i} )  &             & (\rm{ii}) &             & (\rm{iii}) 
             &             & (\rm{iv}) &             & (\rm{v}) 
\label{eq:changeU}
\end{array}
\end{equation}
where $U_i$ ( $i=1,2,3$ ) are defined as
\begin{itemize}
\item
Single-well potential : $U_1(x) = \frac{x^4}{4}$,
\item
Double-well potential : $U_2(x) = \frac{x^4}{4} -\frac{A x^2}{2}
+ \frac{A^2}{4}$,
\item
Bias potential : $U_3(x) = -Bx$.
\end{itemize}
Corresponding to the initial state ``0'' or ``1'', an RTO operation is
expressed as in Fig. 1(a) or 2(b), respectively. 
We have introduced the ``neutral'' state,  ``n'' 
which was not described explicitly by Landauer.
The states ``1'' and ``0'' will be called also the ``memory'' state, ``m''.
The RTO operation is then expressed as a combination of two processes

(A)~~~ (i) $\rightarrow$ (ii) : Erase ``1'' \quad
or \quad 
(A$^{\prime}$)~~ (i$^{\prime}$) $\rightarrow$ (ii) : Erase  ``0'',

(B)~~~ (ii) $\rightarrow$ (iii)  $\rightarrow$ (iv)  
$\rightarrow$ (v) : Record ``1''.

\noindent
Process (B) can be reversible, whereas (A) and (A$^{\prime}$) are 
irreversible because the information has been erased in state ``n''.

Simulating the above processes, we shall answer to questions q1 $\sim$ q4
addressed in the previous section.
 
\vspace{6cm}

\begin{center}
{\small Fig. 1. RTO operation. (a) (A) $+$ (B). (b) (A$^{\prime}$) $+$ (B).}
\end{center}
\noindent


%

\section*{
3. Simulation of a binary device
}

We consider a particle in a 
time-dependent potential described as
the following Langevin equation:
\begin{equation}
m \ddot x = -\gamma \dot x -\frac{\partial U(x,t)}{\partial x}
+ R(t).
\label{eq:langevin}
\end{equation}
The random force $R(t)$ is a white Gaussian noise satisfying
\begin{equation}
\left<R(t) R(t^{\prime})  \right> = 2 \gamma kT, 
\end{equation}
where $T$ is the temperature of the environment.

For example, the process (i) $\rightarrow$ (ii) in 
expression (\ref{eq:changeU}) 
which accompanies the change
of the potential from
$U(x, t_1) = U_2(x)$ to $U(x, t_1+T_1)=U_1(x)$ is described as
\begin{equation}
U(x,t)= \frac{x^4}{4} - \frac{a_-(t-t_1) A x^2}{2}
+ \frac{\{a_-(t-t_1) A\}^2}{4}, 
\end{equation}
where ``switch-on'' function $a_+(t)$ and ``switch-off'' function $a_-(t)$
are defined by
\begin{equation}
a_{\pm}(t) = \left\{
   \begin{array}{ll}
    \displaystyle{\frac{1}{2}}\left( 1 \mp 1\right),
 & \quad t \le 0,\\
\noalign{\vskip0.2cm}
    \displaystyle{\frac{1}{2}}\left( 1 \mp 
\cos [\pi t/T_1],
\right) & \quad 0 \le t \le T_1,\\
\noalign{\vskip0.2cm}
    \displaystyle{\frac{1}{2}}\left( 1 \pm 1 \right),
& \quad t \ge T_1.
\end{array} \right. 
\label{eq:switch}
\end{equation}
%
Similarly, the process 
(ii) $\rightarrow$ (iii) 
in expression (\ref{eq:changeU}) can 
be carried out by changing the potential as
\begin{equation}
U(x,t) = \frac{x^4}{4} 
- a_+(t-t_2) B x.
\end{equation}

The heat generation during a time interval 
$t$ in the environment is given by
\begin{equation}
Q(t) = \int_0^{t} 
\left(\gamma \dot x - R(t^{\prime}) \right) 
\dot x dt^{\prime}. 
\label{eq:q}
\end{equation}
The energy consumption, i.e., the work done by an external agent
is\cite{s97} 
\begin{equation}
W(t) = \int_0^{t}
\frac{\partial U(x, t^{\prime}) }
{\partial t^{\prime}} dt^{\prime}.
\label{eq:w}
\end{equation}
When a process is reversible, 
the entropy of the device changes by the amount

\begin{equation}
\Delta S = - \frac{Q}{T},
\label{eq:s}
\end{equation}
corresponding to the heat generation $Q$ in the environment.

Integrating eq. (\ref{eq:langevin}) numerically, we 
can estimate the heat generation, 
the energy consumption and the entropy change in each process of
computations.

\section*{4. Numerical results and thermodynamical consideration}

We simulated processes (B) $+$ (A) under a constant temperature. 
In order for the system to come thermal 
equilibrium, the potential was kept constant for a time interval $T_0$ 
before and 
after every procedure of changing potential (which takes a time $T_1$, 
see eq. (\ref{eq:switch})). Parameters are fixed as $A=7.0$, 
$B=4.0$, $\gamma=4.0$ and $T_0 =100$. 
$m$, 
$k$ and 
$T$ are  chosen as unity.
For the computer not to work erroneously, it is important that the central
hill of $U_2(x)$, $A^2/4$, must be high enough in comparison with
the thermal energy $k T$ so that the particle in state ``m'' can never
jump over the hill during an ordinary time scale of computations. 
 
After one cycle (B) $+$ (A), both the net work, $W$, done on the device by
the agent (= the energy consumption) and the net heat generation, $Q$, in 
the environment turned out to be almost equal. 
That is consistent with the energy conservation because the device 
returns to the initial state after one cycle.

The processes should tend to be quasi-static as $T_1$ increases.
We examined $T_1$-dependence of $W$ and $Q$ and found that both 
approach to $\sim kT \ln 2= 0.693 kT$ (after one cycle) as 
$T_1 \rightarrow \infty$. This result confirms Landauer's conclusion.

Let us see in detail what happens in each process
to answer to questions q1 $\sim$ q4. 
In Table 1, numbers in lines 1) and 2) are results of our 
simulation, whereas lines 3) $\sim$ 7) are derived from these results based
on thermodynamical consideration. Line 5) represents a net entropy change
of the device, which is the sum of 3) and 4), the former is the entropy 
change corresponding to the heat generation and the latter is the entropy
production. The entropy production is nonzero only in the irreversible
deleting process (A). Line 6) is the entropy change in the environment, which 
directly relates to the heat generation $Q$.

\begin{table}[hbtp]
\caption[I]
{The heat generation $Q$, the work $W$ and the entropy change 
$\Delta S$ in each process. ``D'' and ``E'' stand for
the device and environment, respectively. ``T'' stands for the total 
system: ``D'' $+$ ``E''.
1) and 2) are results of the simulation. Each value is the sample average 
for 100 independent runs. $T_1=1000$. For other parameters, see the text.  
}
\begin{center}
\begin{tabular}{p{1cm}ccccc} \hline
\multicolumn{2}{|c|}{} & 
\multicolumn{1}{c|}{(B)~reversible} & 
\multicolumn{1}{c|}{(A)~irreversible} & 
\multicolumn{1}{c|}{(B) $+$ (A)} &
\multicolumn{1}{c|}{} \\ 
\multicolumn{2}{|c|}{} & 
\multicolumn{1}{c|}{Recording~  ``1''} &
\multicolumn{1}{c|}{Deleting~  ``1''} &
\multicolumn{1}{c|}{One cycle} &
\multicolumn{1}{c|}{} \\
\hline
\multicolumn{2}{|c|}{$Q$ in ``E''}  & 
\multicolumn{1}{c|}{$Q_{\rm r} = 1.07$ ~$\ast$} & 
\multicolumn{1}{c|}{$Q_{\rm d} = -0.36$ ~$\ast \ast$} & 
\multicolumn{1}{c|}{$0.71$} & 
\multicolumn{1}{c|}{1)} \\
\hline
\multicolumn{2}{|c|}{$W$ done on ``D''} &
\multicolumn{1}{c|}{$W_{\rm r} = 1.35$ ~$\diamond$} & 
\multicolumn{1}{c|}{$W_{\rm d} = -0.64$ ~$\diamond \diamond$} & 
\multicolumn{1}{c|}{$0.71$} & 
\multicolumn{1}{c|}{2)}\\
\hline
\hline
\multicolumn{1}{|c|}{} & 
\multicolumn{1}{|c|}{
$\begin{array}{c}
\Delta S \\
{\rm corresponding~ to}~ Q
\end{array}$} &
\multicolumn{1}{c|}{$-Q_{\rm r}/T \stackrel{\ast}{=} -1.07$} &
\multicolumn{1}{c|}{$-Q_{\rm d}/T \stackrel{\ast \ast}{=} 0.36$} &
\multicolumn{1}{c|}{$-0.71$} &
\multicolumn{1}{c|}{
$\begin{array}{c}
3)
\end{array}$} \\
\cline{2-6}
\multicolumn{1}{|c|}{} & 
\multicolumn{1}{|c|}{
$\begin{array}{c}
{\rm entropy} \\
{\rm production}
\end{array}$} &
\multicolumn{1}{c|}{$0$ {\rm (reversible)}} &
\multicolumn{1}{c|}{$0.71$} &
\multicolumn{1}{c|}{$0.71$} &
\multicolumn{1}{c|}{
$\begin{array}{c}
4)
\end{array}$} \\
\cline{2-6}
\multicolumn{1}{|c|}{\raisebox{5ex}[0pt]{``D''}} &
\multicolumn{1}{c|}{
$\begin{array}{c}
\Delta S \\
= 3) + 4)
\end{array}$} &
\multicolumn{1}{c|}{
$\begin{array}{c}
S_1-S_{\rm n} = -Q_{\rm r}/T \\
= -1.07 ~\dagger
\end{array}$} &
\multicolumn{1}{c|}{$S_{\rm n}-S_1 \stackrel{\dagger}{=}1.07$} &
\multicolumn{1}{c|}{$S_{\rm n}-S_{\rm n}=0$} &
\multicolumn{1}{c|}{
$\begin{array}{c}
5) 
\end{array}$} \\
\hline
\multicolumn{1}{|c|}{``E''} &
\multicolumn{1}{c|}{
$\begin{array}{c}
\Delta S\\
{\rm corresponding ~to}~ Q
\end{array}$} &
\multicolumn{1}{c|}{$Q_{\rm r}/T \stackrel{\ast}{=}1.07$} &
\multicolumn{1}{c|}{$Q_{\rm d}/T \stackrel{\ast \ast}{=}-0.36$} &
\multicolumn{1}{c|}{$0.71$} &
\multicolumn{1}{c|}{
$\begin{array}{c}
6)
\end{array}$} \\
\hline
\multicolumn{1}{|c|}{``T''} &
\multicolumn{1}{c|}{
$\begin{array}{c}
\Delta S_{\rm total}\\ 
= 5) + 6)
\end{array}$} &
\multicolumn{1}{c|}{$0$ (reversible)} &
\multicolumn{1}{c|}{$0.71$} &
\multicolumn{1}{c|}{$0.71$} &
\multicolumn{1}{c|}{
$\begin{array}{c}
7) 
\end{array}$} \\
\hline
\end{tabular}
\end{center}
\end{table}
Table 1 explicitly answers to questions q1 $\sim$ q3.
The point is that the energy is consumed and the heat is generated in the
$reversible$ writing process, while the entropy production occurs in the
$irreversible$ deleting process.

Question q4 is answered from the following postulates:

(I) The entropy $S_{\rm m}$ of a recorded 
state ``m'' (m $=$ 1 or 0)
should be estimated from a half of volume of the region in the phase space,
i.e., $0 \le x < \infty$ for ``1''.
This corresponds to the fact that state ``m'' is non-ergodic.

(II) For a deleting process  (``1''$\rightarrow$``n'' or 
``0''$\rightarrow$``n''), the entropy change of the device is the sum of 
two terms, one is $-Q_{\rm d}/T$ corresponding to the heat
generation (in the environment) and the other is an extra increase of 
entropy, $k \ln 2$. Namely,
\begin{equation}
S_{\rm n} - S_{\rm m}= -Q_{\rm d}/T +k \ln 2.
\label{eq:snsm}
\end{equation}

The above postulates are in accordance with the present simulation as we will
show below. We use the general expression for entropy:   
\begin{equation}
S_{\rm n}-S_{\rm m} = \left( \left< E \right>_{\rm n}- 
\left< E \right>_{\rm m} \right)/T + k \ln \left[Z_{\rm n}/ Z_{\rm m} \right].
\label{eq:snsm2}
\end{equation}
The first term, arising from the energy difference, is estimated from 
quadrature as
\begin{equation}
\left( \left< E \right>_{\rm n}- 
\left< E \right>_{\rm m} \right)/T = k/4 - 0.519 k = -0.269,
\label{eq:enem}
\end{equation}
where $U_1(x)$ and $U_2(x)$ are substituted for the potential.
We apply Postulate (I) to estimate the ratio of the partition functions as 
\begin{equation}
Z_{\rm n}/ Z_{\rm m} = \int_{-\infty}^{\infty}e^{-U_1(x)/kT}dx
/ \int_{0}^{\infty}e^{-U_2(x)/kT}dx.
\label{eq:znzm}
\end{equation}
Note that the lower limit of the integral in $Z_{\rm m}$ is $x=0$ 
instead of $-\infty$. The second term of eq. (\ref{eq:snsm2}) is thus 
obtained as $k \ln \left[Z_{\rm n}/ Z_{\rm m} \right] = 1.325$, 
which yields
\begin{equation}
S_{\rm n}-S_{\rm m} 
= 1.056.
\label{eq:snsm3}
\end{equation}
 
Process (B) is reversible, so that the heat generation is given by
$Q_{\rm r}=$ $T \left(S_{\rm n}-S_{\rm m}\right)$ $= 1.056$, 
which agrees well with the simulation (see mark $\ast$ in Table 1).
From the energy conservation, the work is obtained from eqs. (\ref{eq:enem}) 
and (\ref{eq:znzm}) as
$W_{\rm r}=$ $-\left( \left< E \right>_{\rm n}- 
\left< E \right>_{\rm m} \right)$ $+ Q_{\rm r} = 
1.325$,
to be compared with mark $\diamond$ in Table 1.

Next we apply Postulate (II) for the irreversible process (A). 
Then we obtain from eqs. (\ref{eq:snsm}) and (\ref{eq:snsm3}) as 
$Q_{\rm d}=-T \left(S_{\rm n}-S_{\rm m}\right) + kT \ln 2
= -0.363$. The work is, again from the energy conservation, 
$W_{\rm d}= 
\left( \left< E \right>_{\rm n}- 
\left< E \right>_{\rm m} \right) + Q_{\rm d} = -0.632 
$. These values agree well with the results of the 
simulation ($\ast \ast$ and $\diamond \diamond$ in Table 1).

\section*{5. Conclusions and open questions}

We have simulated process (A): ``1''$\rightarrow$``n''.
The same results for $Q_{\rm d}$ and $W_{\rm d}$ in Table 1
should be achieved for process (A$^{\prime}$): ``0''$\rightarrow$``n''
because of the symmetry of the potential.
Also recording ``0'' 
i.e., process (B$^{\prime}$): ``n''$\rightarrow$``0'' yields the same results
for $Q_{\rm r}$ and $W_{\rm r}$ as 
(B)\renewcommand{\thefootnote}{\arabic{footnote}}
\setcounter{footnote}{0}\footnote{Furthermore, 
the order of processes does not affect these results.
For example, (B) $after$ (A) will lead to the same 
$Q_{\rm r}$ and $W_{\rm r}$ as (B) $before$ (A).}.~ 
The entropy in state ``0'' is therefore equal to that in ``1''.
In contrast, the entropy in the deleted state ``n'' is larger as shown in
Table 1 or eq. (\ref{eq:snsm3}).

Now, let us consider a computer composed of $N$ binary devices. After 
a series of computations, each device realizes state ``1'' or ``0'' at random
as
\begin{equation}
0, 1, 0, 1, 1, 0, 1, 0, 0, 1, 0, 1, 1, \cdots 0.
\label{eq:initial}
\end{equation}
Applying RTO on each device in the above, we obtain
\begin{equation}
1, 1, 1, 1, 1, 1, 1, 1, 1, 1, 1, 1, 1, \cdots 1.
\label{eq:final}
\end{equation}
Devices may be in ``1'' or ``0'' before RTO and are
definitely in ``1'' after RTO, but the entropies before and after RTO
are the same in each device. 
If the entropy is an additive quantity, which we admit, the random state
(\ref{eq:initial}) has the same entropy as the ordered state (\ref{eq:final}).
This is one of our conclusions.

It has been widely believed that randomness contributes to entropy.
If so, state (\ref{eq:initial}) would have $N k \ln 2$ larger entropy 
than (\ref{eq:final}). 
Our conclusion contradicts with this traditional view.
The contradiction stems from non-ergodicity of the computer devices.
Even though each device after computations 
seems to have taken ``1'' or ``0'' randomly,
the particles in state ``1'' can never jump into
the side ``0''. Otherwise the computer may make errors!
Exactly the same situation, we believe, occurs in frozen systems like ice.
If the temperature is low enough, each proton in the ice cannot 
jump over the potential hill between two oxygen atoms. 
Thus the system is non-ergodic and the entropy
is the same as the ordered state although the configuration of $N$ protons 
is completely random. 
Therefore we claim that the residual entropy $N k \ln 2$ 
does not exist in the ice despite of the $2^N$-fold degeneracy.
This is another conclusion of ours.

Are our conclusions wrong? We hope not! 
At least we could say that
there are several unsettled problems in thermodynamics when 
non-ergodic systems like computers are involved, which 
were not so important in Boltzmann's time.

\end{document}